\documentclass[twoside,symmetric]{tufte-handout}
\usepackage{hyperref}
\usepackage{mathrsfs}
\usepackage{bibentry}
\usepackage{eufrak}
\usepackage{enumerate}
\usepackage[normalem]{ulem}

\renewcommand{\today}{\number\day \space%
\ifcase \month \or January\or February\or March\or April\or May%
\or June\or July\or August\or September\or October\or November\or December\fi \space%
\number \year} 
\renewcommand{\doi}[1]{\textsc{doi}: \href{http://dx.doi.org/#1}{\nolinkurl{#1}}}
\newcommand{\gev}{\hbox{ GeV}}

\newcommand{\lum}{\hbox{ cm}^{-2}\hbox{ s}^{-1}}

\title{In Leon's company, it seemed\\ that anything might be possible}

\author[Chris Quigg]{Chris Quigg\hfill{\small email:quigg@fnal.gov~~~~~~}\\
{\normalsize Theoretical Physics Department\\ Fermi National Accelerator 
Laboratory\\ P.O. Box 500, Batavia, Illinois 60510 USA}}
\date{{Revised 3 June 2022}\hfill \texttt{\hbox{\small FERMILAB-PUB-20-001-T~~~~}}} 

\usepackage{graphicx} 
  \setkeys{Gin}{width=\linewidth,totalheight=\textheight,keepaspectratio}
  \graphicspath{{graphics/}} 
\usepackage{amsmath}  
\usepackage{booktabs} 
\usepackage{units}    
\usepackage{multicol} 
\usepackage{lipsum}   
\usepackage{fancyvrb} 
  \fvset{fontsize=\normalsize}

\definecolor{IITred}{rgb}{0.5,0.05,0.05}
\definecolor{IITblue}{rgb}{0.05,0.05,0.8}

\begin{document}

\maketitle

\begin{abstract}
\noindent
Memorial sessions celebrated Leon Lederman, Helen Edwards, and Burton Richter at the April 2019 Meeting of the American Physical Society in Denver. 
In the session entitled \emph{Honoring Leon Lederman,} Sally Dawson gave an overview of Leon's scientific career, Marge Bardeen reviewed his work in science education, and I spoke of his years as Fermilab Director. Presentation materials for all three sessions are archived at \url{http://j.mp/31mNkHA}. This essay is drawn from my lecture; my slides can be found at \url{https://bit.ly/2QJ7Jmw}.
\begin{marginfigure}[-96pt]
\centering
\includegraphics[width=0.55\columnwidth]{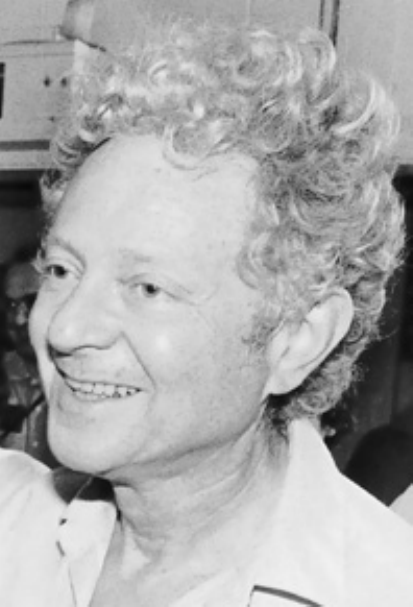}
\break{Leon Lederman in 1983. \\ (Fermilab Creative Services)}
\end{marginfigure}
\end{abstract}

\noindent
\newthought{In 1978, Leon Lederman put aside a promising career} in experimental physics that had spanned three decades 
to accept the appointment as Director of Fermilab---the Enrico Fermi National Accelerator Laboratory. 

Leon was tied to Fermilab even before it existed. He had discovered early in life that if he took himself seriously, others  were likely to take him seriously as well. As the community pondered a new accelerator in the 100-GeV and above range, Leon delivered a manifesto for what he called a \emph{Truly National Laboratory.}\footnote[][-42pt]{L.~M.~Lederman,  ``The Truly National Laboratory (TNL),'' in \emph{Super-High-Energy Summer Study,} Brookhaven Report
No. BNL-AADD-6 (1963), pp. 8--11.}

He gave voice to a complaint of university users that they had not been treated justly at Berkeley and Brookhaven, where the very powerful in-house groups would---according to Leon---consume all the prime beam time for themselves, and only then distribute scraps among the sorry supplicants from Columbia University and other institutions. He enunciated the notion of a TNL (a play on Brookhaven's BNL) that would privilege visiting experimenters. A capable in-house group would maintain the accelerator and could participate in experiments, but the new institution should  be conceived as a paradise for  university users. And, he argued, only in that way could the Truly National Laboratory succeed.\begin{marginfigure}[12pt]
{\includegraphics[width=\columnwidth]{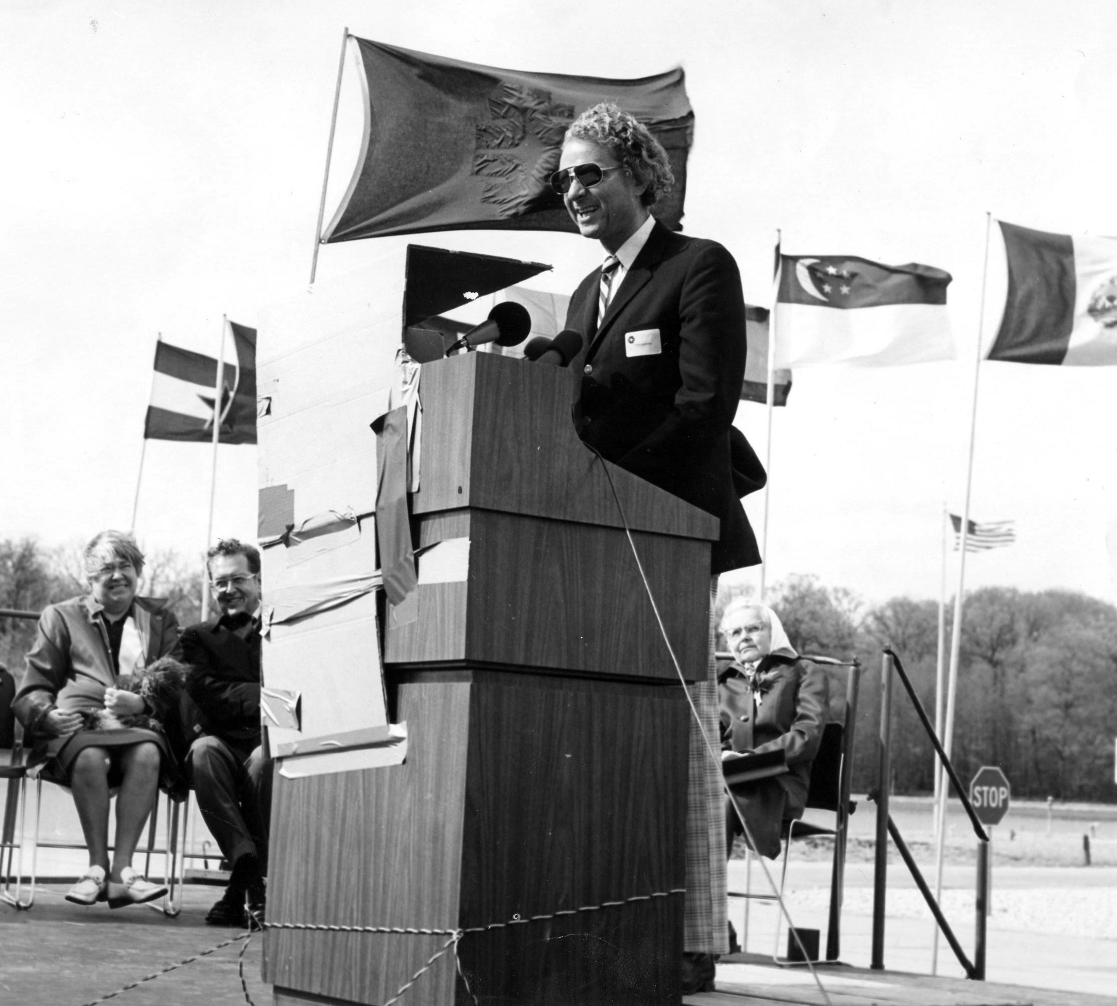}}
\caption{Leon speaking on behalf of users at the Fermilab dedication. (Fermilab Creative Services, 1974)}\label{fig:ded}
\end{marginfigure}

This history contributed to Leon's selection to give the Users' address at the windblown dedication of Fermilab in 1974. In the 
photograph at right you will see Leon (Figure~\ref{fig:ded}) giving his address---impeccably coiffed, I have to say. In the pioneer spirit of Fermilab, the podium is bedecked with a certain amount of cardboard and twine. I don't see any sealing wax, but  duct tape is prominent \ldots \break This illustrates Bob Wilson's dictum that ``Money and effort that would go into an overly conservative design might better be used elsewhere \ldots A major component that works reliably right off the bat is, in one sense, a failure---it is over-designed.\footnote[][-12pt]{R. R. Wilson, ``Sanctimonious Memo \#137,'' July 11, 1969.}

\newthought{Leon would aver in later years} that he had never wanted to be a department chair or lab director. ``Being a professor at a university is the best invention of Western civilization,'' he rhapsodized.  There's where you have power, you have freedom, you can do anything you want.  \ldots Who wants to be a director where you are not free to do anything, everyone is watching you? God help you if you fall asleep, which you often do at seminars, everyone notices and puts it down.''\footnote{Kate Metropolis interview with LML, July 11, 1991.} (Who among us ever saw Leon asleep in a seminar?)

\newthought{His first task as new Director} was to secure a future for the laboratory. A great laboratory has a time horizon of about ten years, in the absence of an idea. Fermilab had completed its first round of many small experiments and several major experiments---Leon and company had just savored the triumph of discovering the Upsilon resonances\footnote{S.~W.~Herb {\it et al.},
  ``Observation of a Dimuon Resonance at 9.5-GeV in 400-GeV Proton-Nucleus Collisions,''
  \href{https://doi.org/10.1103/PhysRevLett.39.252}{Phys.\ Rev.\ Lett.\  {\bf 39}, 252 (1977).}
  }
 and thus the fifth quark, $b$---but it was time  to move on. CERN was coming on strong with its Super Proton Synchrotron, and plans for the S$\bar{p}p$S Collider were taking form, and so Fermilab had to rise to the challenge.

The laboratory had a reputation at the time for having many balls in the air---too many, in the opinion of critics. This is slightly unjust, but it had some basis in fact. A visionary paper from Bob Wilson\footnote[][-6pt]{R.~R.~Wilson,
  ``Fantasies of Future Fermilab Facilities,'' \href{https://doi.org/10.1103/RevModPhys.51.259}{Rev.\ Mod.\ Phys.\  {\bf 51}, 259 (1979).}
  }
 described a profusion of ideas on his mind and on the minds of others. In time, some of them were realized at Fermilab or elsewhere, but the plenitude of possibilities did contribute to the perception that Fermilab couldn't make up its mind about what to do. Leon would have to rally a community with a common commitment.

{Wilson had long dreamed} of building a superconducting synchrotron to supplement the Main Ring. Now, I was young and hadn't yet become acquainted with doubt, so it was quite clear to me what needed to be done:  build the superconducting accelerator and make a proton--antiproton collider out of it. Perhaps Leon had the same insight, but he knew a little more about life. He understood that it wasn't enough merely to announce what was  obviously the right course of action, you had to attend to many constituencies. He took two important steps. While Leon liked to trust his hunches---and he had many---he was careful to verify them. So first he convened Boyce McDaniel, Burt Richter, and Matt Sands, anointing them as the Three Wise Men---to assure him that the superconducting machine could, in fact, be realized. Then he  listened to many ideas. 

\newthought{Leon had a gift for  stagecraft,} so on Armistice Day in 1978 he convened a gathering at which the proponents of colliding the Booster on the Main Ring, the Main Ring on the Tevatron, this little ring on that big ring, protons on protons, protons on antiprotons, all came together and presented their plans to each other and to a jury of their peers. No one could reasonably say, ``You never gave my idea a chance.''  It was clear by the end of that day that Fermilab should undertake something  impressively ambitious, rather than modest and limited---even if that meant giving up any hope of competing with CERN to discover the $W$ and $Z$ bosons, force carriers of the weak interactions.

\begin{marginfigure}[-180pt]
{\includegraphics[width=\columnwidth]{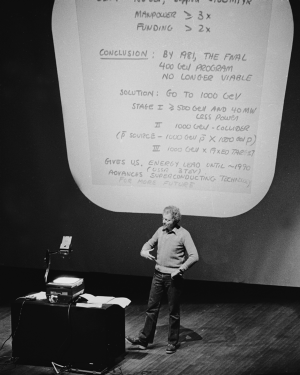}}
\caption{Onward and Upward! \break(Fermilab Creative Services, 1980)}\label{fig:verdict}
\end{marginfigure}

Leon pronounced the verdict (see his transparency in Figure~\ref{fig:verdict}), noting that the Fermilab program would no longer be competitive starting around 1981, that we had to move on, and that advancing toward higher energy was an essential part of a glorious future for science and the laboratory. The Armistice Day consensus foresaw two different applications of the Tevatron, or the Energy Doubler/Saver, as it was called at the time. The fixed-target Tevatron program would be the first step, giving new life to many existing experimental efforts; a colliding beams program featuring proton--antiproton collisions would follow later. 

 \newthought{With The Plan in hand,} Leon went off to the High Energy Physics Advisory Panel in Washington in the early spring with me in tow. He talked for fifty-five minutes about how Fermilab had made decisions, set clear priorities. We would, he vowed, build the superconducting accelerator and mount a reinvigorated fixed-target program, bigger and better than ever. Then I had five minutes as a kid at the end of the hour, in which I showed a single slide of parton luminosities and talked about the $\bar{p}p$ collider that would follow in good time.

A few days later, late on a Friday afternoon, Leon got a phone call from Washington saying, ``We can sell the collider. Get a plan in by Monday!'' (for a price the caller named) and so Leon, Helen Edwards, Alvin Tollestrup, and other gifted people  worked through the weekend to put together a sort-of-plausible plan for a Tevatron Collider. I think we submitted it on Tuesday, requesting more money than they told us we were allowed to have. And it was actually approved, after appropriate scrutiny. Washington's enthusiasm for the collider shuffled our Plan only slightly---the collider project was named Tevatron I and the fixed-target project Tevatron II. Keeping our promises was no small task!

\begin{marginfigure}[-112pt]\includegraphics[width=\columnwidth]{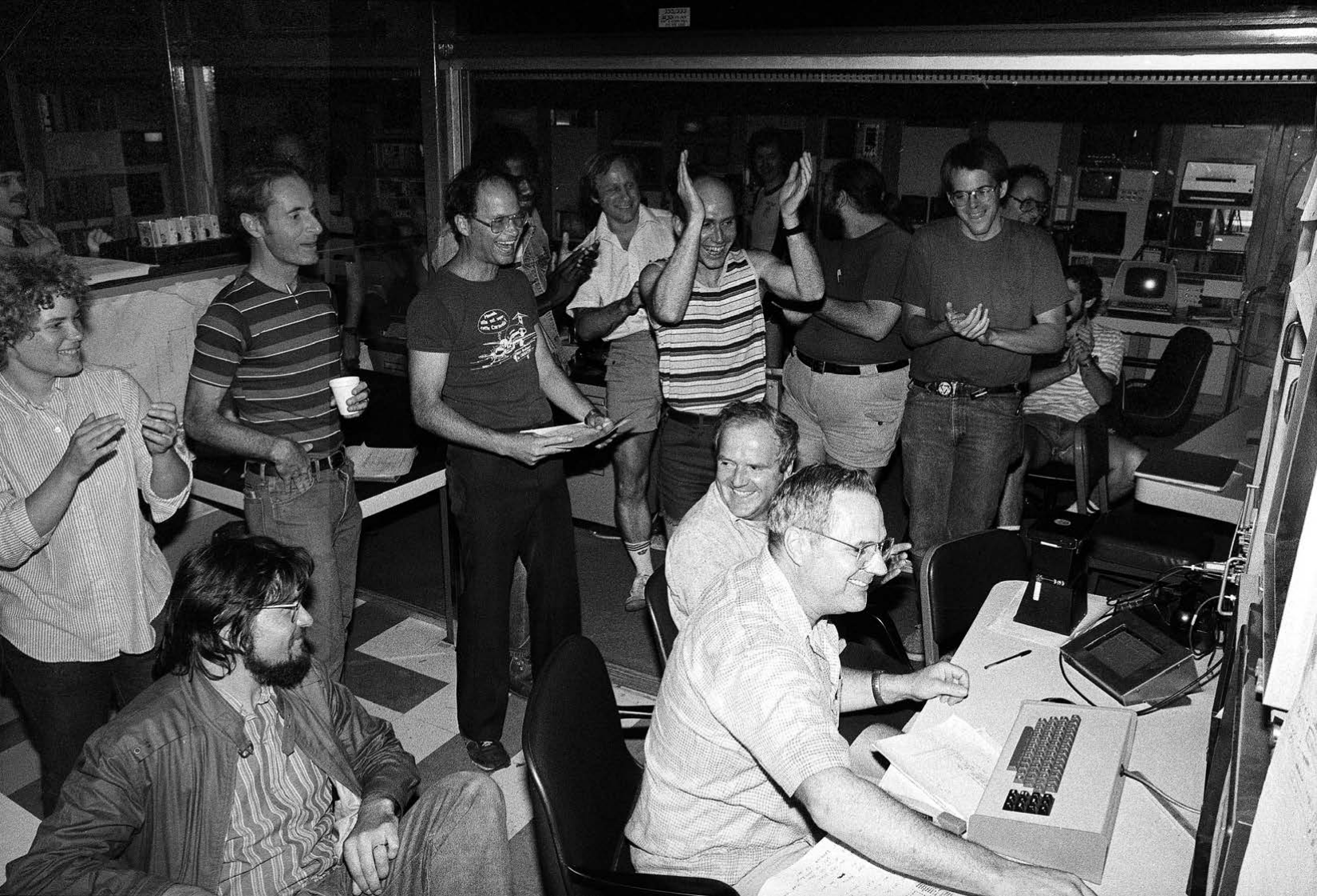}\caption{The Tevatron reaches $512\gev$!\break (Fermilab Creative Services, 1983)}\label{fig:512gev}\end{marginfigure}

Heroic efforts by many completed the design of the machine, the development and mass production of the superconducting magnets, and the refinement of control and quench-protection systems. The Tevatron was installed in the Main Ring tunnel in free space below the Main Ring magnets. Figure~\ref{fig:512gev} shows the mood in the control room at the moment on July 3, 1983, that a proton beam circulated at 512 GeV, then a world record, showing that a superconducting synchrotron was practical. Particle physics instantly came alive with new possibilities. The technological achievement resulted in rewards for many people, including Helen, Dick Lundy, Rich Orr, and Alvin, each of whom received the National Medal of Technology. 

\begin{marginfigure}[-24pt]\includegraphics[width=\columnwidth]{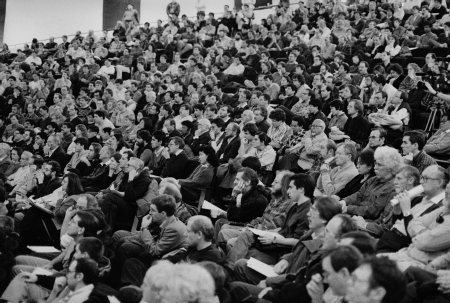}\caption{Top quark announcement \break(Fermilab Creative Services, 1995)}\label{fig:topquark}\end{marginfigure}

Creating the collider required further enormous effort, beginning with the superb antiproton source built by the team led by John Peoples. A dozen years into the life of the Tevatron we cheered the discovery of the top quark  by the CDF and D0 Collaborations. Figure~\ref{fig:topquark} shows Ramsey Auditorium at capacity for the discovery seminars. It was a great  day for Fermilab and for physics. 

\newthought{Leon had set about enriching the lab's scientific life} from his first moments as Director-Designate.The day he took the job, he telephoned me---I am sure this was one call among many---to say, ``We'' (not You, or I, but We) ``have to keep building up theory at Fermilab.'' He was unfailingly supportive. In particular, he convinced James Bjorken to come and spend a decade at the laboratory, taking on a generalized role as a sort of Chief Scientist. Along the way, Bj took an interest in accelerator physics, winning the Wilson Prize for studies of intrabeam scattering. Nurturing theoretical physics was one of many things that Leon did to make science even more central to the lab than it had been before.

David Schramm whispered into Leon's ear that it was time to get cosmology together with particle physics at the lab. Leon again trusted his hunch but verified. The way we verified was by having a year-long series of seminars in which established figures would visit and lecture so the lab staff could get some footing in this subject.  Leon could pick their brains about the right way to go about incorporating cosmology. We began with Jim Peebles and then went on through a series of other experts. By 1984, the new NASA/Fermilab Theoretcal Astrophysics Center announced itself with a memorable Inner Space / Outer Space conference.\footnote{E.~W.~Kolb, et al. (Eds.),
 \emph{Inner Space / Outer Space: The Interface Between Cosmology And Particle Physics, May 2--5, 1984} (University of Chicago Press, Chicago, 1986).} 
 
 Leon instigated a Scientific Advisory Group that met first thing on Monday mornings. It was a vehicle not only for him to hear from other people, but for them to take Leon's thoughts and concerns back to the rest of the laboratory. Often there were real scientific discussions in those meetings, including debates about scientific strategy and priorities. Sometimes there was a therapeutic element as the director vented about the indignities that had been visited on him by the agency or rival lab directors. 

Leon followed up with a Junior SAG to involve more people. Younger colleagues have told me that being named to that group made them feel that they were really part of the laboratory and had a contribution to make to the institution. Leon had a gift for including people and motivating them to invest in the lab's future.

\newthought{Leon worked hard to enhance the quality of life} at Fermilab. Having the perspective of a university user as well as Director, he was motivated to improve the Fermilab experience for everyone---users, families, and lab staff broadly understood. An important early contribution was changing the Users Center into a place that was welcoming for many people. He found Tita Jensen, an accomplished chef and a brilliant hostess as well, who made Chez Leon, as the restaurant soon came to be called, a meeting place for Fermilab for several decades. 

\begin{marginfigure}[12pt]\includegraphics[width=\columnwidth]{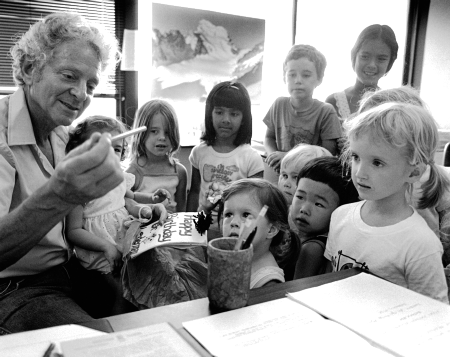}\caption{Leon at the Children's Center \break (Fermilab Creative Services, 1982)}\label{fig:bidet}\end{marginfigure}

With the  help of many Fermilab colleagues and the local DOE office (who moved mountains and made many miracles) Leon created the first Children's Center (daycare center) in a national laboratory. In Figure~\ref{fig:bidet}, he is celebrating one of his birthdays with the children.

Leon continued to boost the lab's intellectual climate. He founded a series of Director's Colloquia featuring scientific luminaries, to emphasize the importance of scientific culture. In addition to physicists doing important contemporary work---we should actually have a few of them back to give talks with the same titles, because the issues are still current---there were celebrated people of an earlier era such as  Viki Weisskopf and Leon's teacher I. I. Rabi,  as well as Leon's contemporaries, including Marvin Minsky. Leon would organize, attend, entertain, and introduce the visitors to young colleagues. These events displayed Leon in his element, and were a great gift to us all.

To lure colleagues out of their offices and into conversation, he invented the Director's Coffee Break, which promised free coffee and cookies every afternoon at 3:30 in the gallery space near his office. He would appear regularly and find his way into this group or that---listening, questioning, performing. Leon's accessibility promoted a sense of unity and purpose.

He was also famous for visiting experiments unannounced. Not all of these visits took place at 2:30 in the morning, but enough to be the stuff of legend. He knew where to find the Main Control Room  and would drop in on the people making the accelerator hum.

\newthought{Leon began to use the Director's bully pulpit} even before taking office. South America had a storied tradition in cosmic-ray physics, but in the late 1970s, particle physics south of the Rio Grande was mostly theoretical in character. Leon wanted to build up experimental groups to participate in research at Fermilab and other accelerator laboratories. With friends and colleagues from the South---including Germinal Cocho (UNAM), Carlos Garc\'ia-Canal (Argentina), and Roberto Salmeron and Jos\'e Leite-Lopes (Brazil)---he organized a symposium to promote the idea. When Leon did something, it was not a halfway effort! 
He brought along a small Academy of his famous friends from the North (many with a pre-existing interest in the South)---Bjorken, Charpak, Feynman, Glashow, Marshak, Moravscik, Richter, Tollestrup, Samios, Panofsky, Dick Taylor, and Bob Wilson---to plan together and impress the local authorities.\footnote[][-3pt]{A short paper, L.~M.~Lederman, ``Fermilab and Latin America,''
  \href{ https://doi.org/10.1063/1.2359388}{AIP Conf.\ Proc.\  {\bf 857}, 15--23 (2006)},  summarized his aspirations.
    }
    
Such efforts helped catalyze a growing intellectual commerce between Fermilab and Latin-American institutions, involving not only physics Ph.D.s and students, but engineers and technicians as well. Leon outlined much grander visions, not all of which came immediately to fruition. In 1984 he proposed that the United States  offer our Latin American neighbors a massive graduate fellowship program in science and engineering. He believed that changing lives could change the relationships among nations, and he made the case that this would not be a giveaway, but a high-reward investment for the USA. Perhaps it is an idea whose time has come?

\begin{marginfigure}[-112pt]\includegraphics[width=\columnwidth]{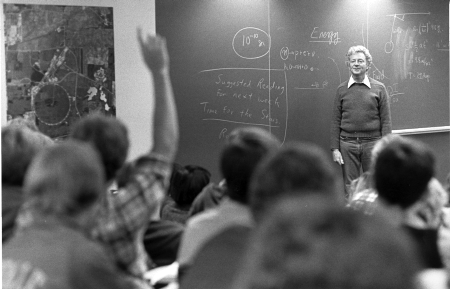}\caption{Leon launching a new session of Saturday Morning Physics. \break (Fermilab Creative Services, 1980)}\label{fig:SMP}\end{marginfigure}

\newthought{Leon thrived on interactions with young people.} I think that one of his life's missions was to nourish their innate curiosity---to help them \emph{not} grow up, as he would have put it. Very early in his time at Fermilab, he enlisted many people in creating Saturday Morning Physics, a program in which high school students would come to the lab for ten Saturday mornings, hear lectures for a couple of hours, then participate in tours or other educational activities. We noticed early on that, because some of the kids weren't old enough to drive, their high school teachers would chauffeur them, giving up ten of their Saturday mornings in a row. Marge Bardeen has described what came of that.\footnote[][-144pt]{M.~Bardeen, ``The Singing Janitor and Other Stories: Leon's Adventures in Science Education,''  talk at APS April Meeting (Denver, 2019) \url{https://bit.ly/2ttsFWG}.} Leon well understood the value of the Director himself showing up to welcome the students, set the tone in the opening lecture, and respond to questions (Figure~\ref{fig:SMP}). 

Sally Dawson has mentioned meeting Leon as a student at the Les Houches Summer School in 1981.\footnote[][-12pt]{S.~Dawson, ``From Neutrinos to $b$s:
The incredible physics contributions of Leon Lederman,'' talk at APS April Meeting (Denver, 2019) \url{https://bit.ly/2ZTLkqX}.} Leon not only appeared at the pig roast, he gave a series of lectures.\footnote[][6pt]{L.~M.~Lederman, ``High Energy Experiments,'' in \emph{Gauge Theories in High Energy Physics,} Les Houches Session XXXVII, edited by Mary K. Gaillard and Raymond Stora (North-Holland, Amsterdam, 1983), pp. 927--863.
} It seems to be harder these days for lab directors to find time to give courses at summer schools, but Leon's availability was a very effective recruiting tool.

Leon and others started a university--Fermilab Ph.D. program in accelerator physics and technology  to introduce accelerator science, which had been propagated mainly in the laboratories, into a greater number of universities than before.

%

\begin{marginfigure}[-42pt]\centerline{\includegraphics[width=0.55\columnwidth]{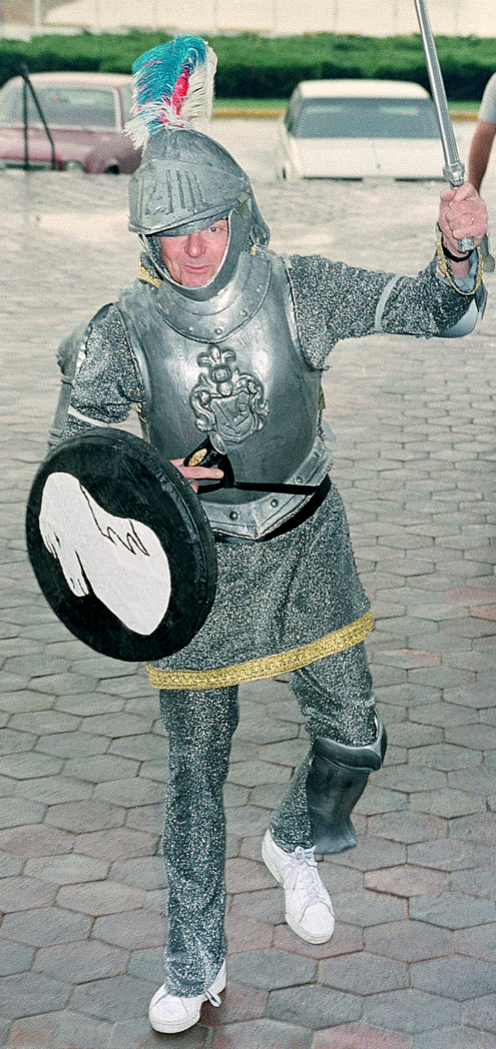}\phantom{MMMM}}\caption{The People's Champion \break (Fermilab Creative Services, 1986)}\label{fig:knight}\end{marginfigure}

\newthought{Leon  brought unprecedented dignity to the office.} Here he is (Figure~\ref{fig:knight}) arriving for a lab-wide celebration as the champion of the people and defender of the faith. I don't know what precisely he was defending us against, but there were plenty of dragons out there. At an annual lab picnic, he took his turn in the dunk tank, where he proved to be an irresistible target for children and also for colleagues with minor scores to settle. He championed Miss Illinois USA 1981, whose costume celebrated Fermilab and the discovery of the Upsilon. That was, alas, not enough to propel her to the title of Miss USA. 

\newthought{Once the Tevatron construction was under way,} Leon began to look over the horizon.  In the spring of 1977, I gave a talk at Brookhaven, just after we had identified a simple argument for the importance of the TeV scale. Leon sat in the front row doodling or maybe taking notes, as I talked about how essential it was to explore the TeV scale, which was far beyond the capabilities of ISABelle, as planned at the time. He looked up at me with a sneer and hissed, ``You guys, you're never satisfied!'' But he did get the message.

Only a short time later, he had found a number of people with stars behind their names who were giving the same advice. He gave a talk at the Snowmass Summer Study in 1982,\footnote{ L.~M.~Lederman,
  ``Fermilab and the Future of HEP,''  in \emph{Snowmass 1982, Proceedings, Elementary Particle Physics and Future Facilities,} edited by R.~Donaldson, H.~R.~Gustafson, and F.~E.~Paige \href{https://inspirehep.net/record/185815/files/C8206282-pg125.PDF}{eConf C {\bf 8206282}, 125--127 (1982).}
  }  advocating the quest for a hadron collider of at least 10 on 10 TeV, maybe 20 on 20 TeV. To make this happen, he said, you have to eliminate pernicious competition by building the new machine away from any of the existing laboratories. The notion of a new laboratory, perhaps in a remote location, became known as the Desertron.

Immediately after the decision in 1983 to push forward with  the Superconducting Super Collider, Leon launched an effort to get people engaged in detector issues, recruiting Bruce Winstein from the University of Chicago as coordinator. Thus was born the Physics at the SSC study group, a roadshow of meetings dedicated to different technologies for detectors.\footnote[][-48pt]{P.~Hale and B.~Winstein, (Eds.)
  ``PSSC Physics at the Superconducting Super Collider Summary Report : A Summary Report of the PSSC Discussion Group Meetings,''
  \href{http://inspirehep.net/record/201623/files/PSSC.pdf}{SSC-PSSC-SUMM.1.}
 }

He brought together essentially all of the Immortals---physicists with Nobel Prizes---and governors of states competing to host the new laboratory, to write statements in support of the SSC.\footnote{  L.~M.~Lederman and C.~Quigg,
  \href{https://inspirehep.net/record/272205/files/AppraisingtheRing.pdf}{\emph{Appraising the Ring: Statements in Support of the Superconducting Super Collider}} (Universities Research Association, Washington, DC, 1988)
  }

\newthought{One of Leon's conditions for taking the job} was that he be able to continue his career as an experimental physicist. He later said that this didn't work out; the job of lab director was just too big. But he did not stop thinking about physics. He joined the Hyper-CP experiment where, I am told,  he made a crucial suggestion about the orientation of an analysis magnet.

Leon wrote a paper\footnote[][-18pt]{ R.~Huson, L.~M.~Lederman and R.~F.~Schwitters,
  ``A Primer on Detectors in High Luminosity Environment,''
  \href{https://lss.fnal.gov/conf/C8206282/pg361.pdf}{eConf C {\bf 8206282}, 361 (1982).}
  } with Russ Huson and Roy Schwitters on the impossibility of doing experiments at luminosities higher than $10^{32}\lum$. This was received in some quarters as an assault on higher-luminosity, lower-energy machines, but it clearly identified the obstacles to high-luminosity experimentation. Should you contemplate giving a course on LHC detectors, it is very instructive to compare the challenges foreseen in this paper with what has been achieved. To see how our colleagues have solved these problems in the intervening time is very illuminating. To his great credit, Leon found time as lab director to pursue such questions.

\begin{marginfigure}[-64pt]\hspace*{2.5ex}{\includegraphics[width=0.7\columnwidth]{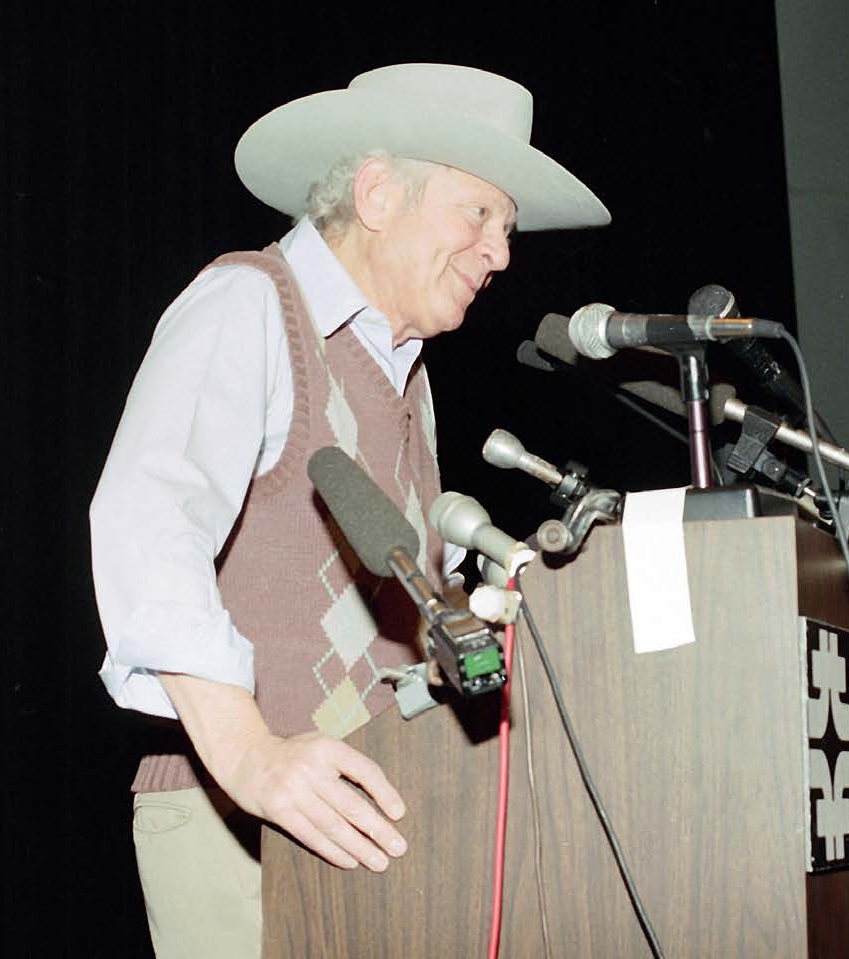}}\caption{Hats off to Waxahachie!\break (Fermilab Creative Services)}\label{fig:Waxa}\end{marginfigure}

\newthought{Toward the end of his mandate,} Leon  answered a call from Stockholm,\footnote[][-24pt]{\href{https://www.nobelprize.org/search/?s=lederman}{Nobel Prize in Physics, 1988,}  ``for the neutrino beam method and the demonstration of the doublet structure of the leptons through the discovery of the muon neutrino.''}  which unleashed a time of joyous festivities at Fermilab. Soon after, on the day after the presidential election in 1988, the Department of Energy announced that it would site the Superconducting Super Collider in Waxahachie,  Texas, not Batavia, Illinois. For Leon, celebration gave way to pastoral counseling. He turned up for an all-hands meeting wearing a full-blown Stetson (Figure~\ref{fig:Waxa}) and sought to reassure his colleagues that Fermilab would endure and thrive. Very quickly, he and his team put together a plan for the Main Injector and the Recycler ring. This is now the principal accelerator installation at Fermilab.

\newthought{Leon passed the Director's baton to John Peoples} after ten years, and had a long time to savor the afterlife. As Director, he had launched the Illinois Mathematics and Science Academy; after stepping down, he spent a lot of time there as Scientist-in-Residence. He taught at the University of Chicago and at the Illinois Institute of Technology. He served as president of the American Association for the Advancement of Science. He was active in the \emph{Bulletin of the Atomic Scientists} and had a wide-ranging influence as a statesman of science. 
He published five books: the infamous \emph{God Particle,} an earlier book with Dave Schramm on the interplay between particle physics and cosmology, and three engaging volumes with my colleague, Chris Hill.\footnote[][-132pt]{L.~M.~Lederman and D. N. Schramm, \emph{From Quarks to the Cosmos} (W.~H.~Freeman, San Francisco, 1989). L.~M.~Lederman with Dick Teresi, \emph{The God Particle} (Houghton Mifflin Harcourt, Boston, 1993). L.~M.~Lederman and C.~T.~Hill,
\emph{Symmetry and the Beautiful Universe} (Prometheus, Amherst, NY, 2004); \emph{Quantum Physics for Poets} (Prometheus, Amherst, NY, 2011); \emph{Beyond the God Particle} (Prometheus, Amherst, NY, 2013).}

\begin{marginfigure}[-36pt]
\includegraphics[width=\columnwidth]{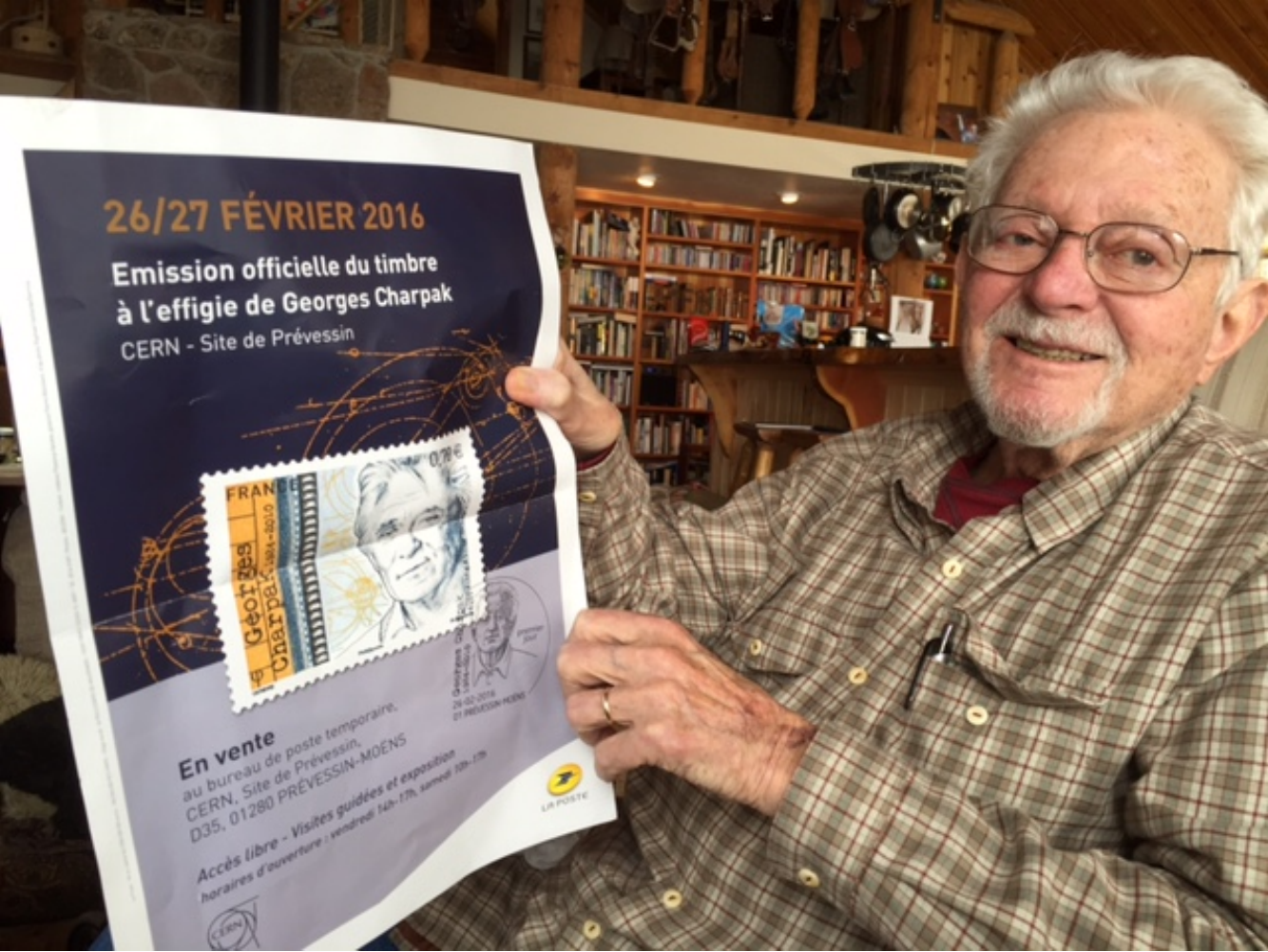}\caption{Leon and Georges. \break(Ellen Lederman photo, 2016)}\label{fig:Charpak}
\end{marginfigure}

Throughout his career, Leon was a magnet for talent. One of the people that he attracted was his great friend, Georges Charpak. After Georges's death, when the French government issued a commemorative stamp in his honor,  our friends at CERN were able to provide Leon with a poster marking the first day of issue (Figure~\ref{fig:Charpak}).

\newthought{Leon  had a prodigious capacity for making us smile.} Almost everyone who encountered him  will have heard many of his stories over and over again. I don't know whether it was sympathy that made us respond to him with such affection, or solidarity with the simple delight he took in performance, or actual amusement, but jokes and shaggy-dog stories were part of what made Leon the great character he was.

\begin{marginfigure}[-192pt]\includegraphics[width=\columnwidth]{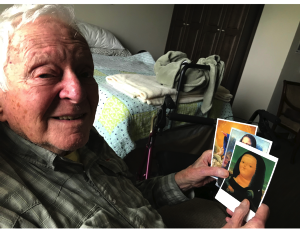}\caption{Leon appreciating fine art.\break (Ellen Lederman photo, 2017)}\label{fig:monas}\end{marginfigure}

Late in his life, when Leon was in cognitive decline,  I learned from his wife Ellen that he took great pleasure in receiving picture postcards. Knowing that he had what used to be called an eye for the ladies, I first sent him a postcard of the Mona Lisa, to which he responded, ``I used to date her!'' This was a flash of the old Leon. Before I heard the reaction, I had sent him a second Mona Lisa, by Fernando Botero, one of three cards he is holding in Figure~\ref{fig:monas}. I told this story to our Administrative Assistant, Olivia Vizcarra, who knew a bit about Leon's persona. She said, ``Send him the Virgin Mary!'' The third card in Leon's hands is an image of the Assumption.\footnote{Guido Reni (1637), \href{https://www.mba-lyon.fr/fiche-oeuvre/lassomption-de-la-vierge}{Mus\'ee des Beaux Arts de Lyon.}} I leave to your imagination what Leon might have said in response.

\newthought{I thank Leon for all that he gave us.} My title sums up the essence of the Lederman years at Fermilab: \emph{In his company, it seemed that anything might be possible.}

\vspace*{36pt}\noindent
For their help in providing archival materials, photographs, and advice, I am grateful to  Reidar Hahn,Valerie Higgins, Karin Kemp, Ellen Lederman, Kate Metropolis, and Karen Seifrid. I thank Young-Kee Kim and Pier Oddone for the invitation to present the lecture on which this essay is based.

\vspace*{24pt}\noindent
This work was supported by Fermi Research Alliance, LLC, under Contract No. DE-AC02-07CH11359 with the U.S. Department of Energy, Office of Science, Office of High Energy Physics.


\end{document}